\documentclass{PoS}
\usepackage{subfigure}

\title{$K$-meson vector and tensor decay constants and $B_{\rm K}$-parameter from $N_f = 2$ tmQCD}

\ShortTitle{$f_{K^*}$, $f_T$ and $B_K$ from tmQCD}

\author{ETM Collaboration}

\author{P.~Dimopoulos, R.~Frezzotti, G.C.~Rossi\\
        Dip. di Fisica, Universit\`a di Roma ``Tor Vergata",\\
        Via della Ricerca Scientifica 1, I-00133 Rome, Italy \\
        E-mail: \email{\{dimopoulos,frezzotti,rossig\}@roma2.infn.it}}

\author{V.~Gimenez\\
        Dep. de Fisica Te\`orica and IFIC, Univ. de Val\`encia-CSIC,\\
        Dr.~Moliner 50, E-46100 Val\`encia, Spain\\
        E-mail: \email{vicente.gimenez@uv.es}}

\author{V. Lubicz\\
         Dip. di Fisica, Universit{\`a} di Roma Tre,\\ 
         Via della Vasca Navale 84, I-00146 Rome, Italy \\
         E-mail: \email{lubicz@fis.uniroma3.it}}

\author{F.~Mescia\\
        INFN-Laboratori Nazionali di Frascati,\\
        Via E.~Fermi 40, I-00044 Frascati, Italy \\
        E-mail: \email{mescia@lnf.infn.it}}

\author{S. Simula\\
         INFN-Roma Tre, c/o Universit{\`a} di Roma Tre,\\ 
         Via della Vasca Navale 84, I-00146 Rome, Italy \\
         E-mail: \email{simula@roma3.infn.it}}

\author{\speaker{A.~Vladikas}\\
        INFN-``Tor Vergata", c/o Dip. di Fisica, Universit\`a di Roma ``Tor Vergata",\\
        Via della Ricerca Scientifica 1, I-00133 Rome, Italy \\
        E-mail: \email{vladikas@roma2.infn.it}}

\abstract{We present work in progress  on the computation of the $K$-meson vector and
tensor decay constants, as well as the $B$-parameter in Kaon oscillations. Our simulations are
performed in a partially quenched setup, with two dynamical (sea) Wilson quark flavours, having
a maximally twisted mass term. Valence quarks are either of the standard or the Osterwalder-Seiler
maximally twisted variety. These two regularizations can  be suitably combined in order to obtain
a $B_{\rm K}$ parameter which is both multiplicatively renormalizable and $O(a)$ improved. 
}

\FullConference{The XXVI International Symposium on Lattice Field Theory\\
		 July 14-19 2008\\
		 Williamsburg, Virginia, USA}

\begin{document}

\section{Introduction: theory and computational setup}

In the last few years, the ETM Collaboration (ETMC) has been carrying out state-of-the-art lattice QCD simulations with $N_f=2$ dynamical flavours (sea quarks) and `lightish" pseudoscalar meson masses (300 MeV $< m_{\rm PS} < $ 550 MeV).
Strangeness clearly enters the game in a partially quenched context. Several physical quantities are currently being analyzed for a few lattice spacings. In the present work we will present {\it preliminary} results on: (i) the vector meson mass $m_{\rm K^*}$, (ii) its decay constant $f_{\rm K^*}$,
(iii) the ratio of tensor to vector decay constants $f_{\rm T} / f_{\rm V}$, computed at the $\rm K^*$ physical mass and (iv) the $B_{\rm K}$ parameter for neutral Kaon oscillations. Other subgroups of the collaboration are working on decay constants in the light and strange quark sector ~\cite{Craig,Cecilia}.

ETMC simulations are performed with the tree-level Symanzik improved gauge action. The $N_f = 2$ sea quark flavours are regularized by the standard Wilson fermion action with a maximally twisted mass term (referred to as ``standard tmQCD case")~\cite{tmQCD1} . This means that the the two light flavours are organized in a flavour doublet $\bar \psi^T = ( \bar u \,\,\, \bar d)$ and the fermion lattice action is given by
\begin{equation}
{\cal L}_{tm}  \,\, = \,\, \bar\psi \, \Big [ D_W \,\, + \,\, i \mu_q \tau^3 \gamma_5 \Big ] \, \psi 
\end{equation}
with $D_W$ denoting the critical Wilson-Dirac operator. This formulation has well known advantages, amongst which: (i) Renormalization properties are much simpler than with standard Wilson quarks, in many cases of interest (e.g. pseudoscalar decay constants, chiral condensate, $B_{\rm K}$...);  (ii) improvement is automatic at maximal twist~\cite{Frezz-Rossi1}. One should note, however that
for Weak Matrix Elements (WMEs) of 4-fermion operators (e.g. $B_{\rm K}$), it is not possible to retain both
automatic improvement (through the standard tmQCD formalism, with all flavours at maximal twist) and multiplicative renormalization for the relevant operator; see refs.~\cite{AlphaBK,PenSinVla}. One way out is provided by the Osterwalder-Seiler (OS) variant of tmQCD, in which valence quarks enter with a distinct action for each flavour and a fully twisted mass term~\cite{Frezz-Rossi2}:
\begin{equation}
{\cal L}_{OS}  \,\, = \,\, \bar\psi_f \, \Big [ D_W \,\, + \,\, i \mu_f \gamma_5 \Big ] \, \psi_f 
\qquad \qquad f \,\, = \,\, u, \, d, \, s\, \cdots
\end{equation}
Suitable combinations of $\mu_f$ signs for each flavour in the above action ensure automatic improvement and multiplicative renormalization for say, $B_{\rm K}$. The OS option is a compromise, since unitarity issues arise at finite lattice spacing when sea and valence flavours are treated differently. Nevertheless, in our partially quenched setup ($N_f = 2$ sea quark flavours and a valence strange quark) compromising unitarity is anyway unavoidable in any regularization.

The ETMC runs are performed at three gauge couplings $\beta$. Here we report work in progress, confining ourselves to the ``master run" at $\beta = 3.90$, corresponding to a lattice spacing of $a \simeq 0.086(1) {\rm fm}$ (i.e. $a^{-1} \simeq 2.3 {\rm GeV}$). The lattice volume is $ V = 24^3 \times 48$. Our ensemble consists of 240 gauge field configurations for the $K^*$-meson mass and decay constants and $200$ configurations for $B_{\rm K}$. The sea quark mass is set at five values
$a \mu =$ 0.0040, 0.0064, 0.0085, 0.0100, 0.0150, corresponding to pseudoscalar meson masses in
the range $300 {\rm MeV} < m_{\rm PS} < 550 {\rm MeV}$. The valence quark masses are set at seven values; those of the sea quarks plus
the values $a \mu= 0.0220, 0.0270$, which are meant to
facilitate the interpolation to the physical strange quark mass. We use the calibration results of previous ETMC
work~\cite{etmc-light,etmc-strange}; e.g. the physical light quark mass is at $a \mu_d = a \mu (m_\pi) = 0.00079$, whereas the physical strange quark mass  is at $a \mu_s = a \mu(m_{\rm K}) = 0.0217(10)$. For $B_{\rm K}$ only, we checked for finite volume effects 
by running also at $V = 32^3 \times 64$ at the lowest quark mass ($a \mu = 0.0040$). 

For the computation of correlation functions, we used stochastic sources of the extended "one-end trick" of refs. \cite{FostMich,NeilMich}. For a concise exposition of the method see also ref. \cite{etmc-long}.

\section{Vector meson masses and decay constants}

The vector and tensor decay constants are defined by the formulae: 
\begin{eqnarray}
\langle 0|{\cal V}_k| V;\lambda \rangle = f_V \epsilon_k^\lambda m_V \qquad \qquad
\langle 0|{\cal T}_{0k}| V;\lambda \rangle = -i f_T \epsilon_k^\lambda m_V
\end{eqnarray}
Vector meson masses and the decay constants $f_V$, $f_T$ are computed from the correlators
(in continuum notation)
$C_{V_kV_k}(t) = \sum_{\vec x, k} < {\cal V}_k(x) \,\, {\cal V^\dagger}_k (0) >$ and
$C_{T_{0k}T_{0k}}(t) = \sum_{\vec x, k} < {\cal T}_{0k}(x) \,\, {\cal T^\dagger}_{0k} (0) >$.
The best estimate of the vector meson effective mass is provided by the correlator
$C_{V_kV_k}(t) $. The estimate from the correlator $C_{T_kT_k}(t)$ is much noisier.
We compute  $f^{T}/f^{V}|_{K^{*}}$  from the large time asymptotic behaviour of the ratio
$[C_{T_{0k}T_{0k}}(t) / C_{V_kV_k}(t) ]^{1/2}$.

As discussed in ref.~\cite{petros}, the correctly normalized vector current in the tmQCD fermion regularization is ${\cal V}_\mu = Z_A A_\mu^{\rm tm}$, while in the OS setup we have
${\cal V}_\mu = Z_V V_\mu^{\rm OS}$. With $\tilde T_{\mu\nu} \equiv \epsilon_{\mu\nu\rho\sigma} T_{\rho\sigma}$,
we similarly have for the tensor density 
${\cal T}_{\mu\nu} = Z_T T_{\mu\nu}^{\rm tm}$ and ${\cal T}_{\mu\nu} = Z_T \tilde T_{\mu\nu}^{\rm OS}$.
We use the RI/MOM~\cite{rimom2} estimates for the (re)normalisation constants, computed in a
tmQCD framework in ref.~\cite{petros}, obtaining $Z_T(a^{-1};\beta = 3.9) = 0.769(4)$,
$Z_A(\beta = 3.9) = 0.771(4)$ and $Z_V(\beta = 3.9) = 0.6104(2)$.
The slight differences between the above values and the ones of ref.~\cite{petros} are due to a revised analysis.

We wish to highlight straightaway the two problems we have encountered: (i) For all sea quark masses, when the valence quark masses are in the lightest range (say, around $\mu_{\rm val} = 0.0040$), the vector meson effective mass has a poor plateau. The situation already improves at the next mass value of our simulation, $\mu_{\rm val} = 0.0064$. Nevertheless, since the signal-to-noise ratio behaves as expected (i.e. it drops like $\exp[ - ( m_{\rm V} - m_{\rm PS} ) t ]$),  the $\rho$-meson mass and decay constant can still be extracted (see results presented in  ref.~\cite{Craig}). (ii) A poor quality vector meson effective mass is  also seen when $\mu_{val} < \mu_{sea}$. This problem is absent in the pseudoscalar channel.

\begin{figure}[!h]
\begin{center}
\subfigure[]{\includegraphics[scale=0.26,angle=-0]{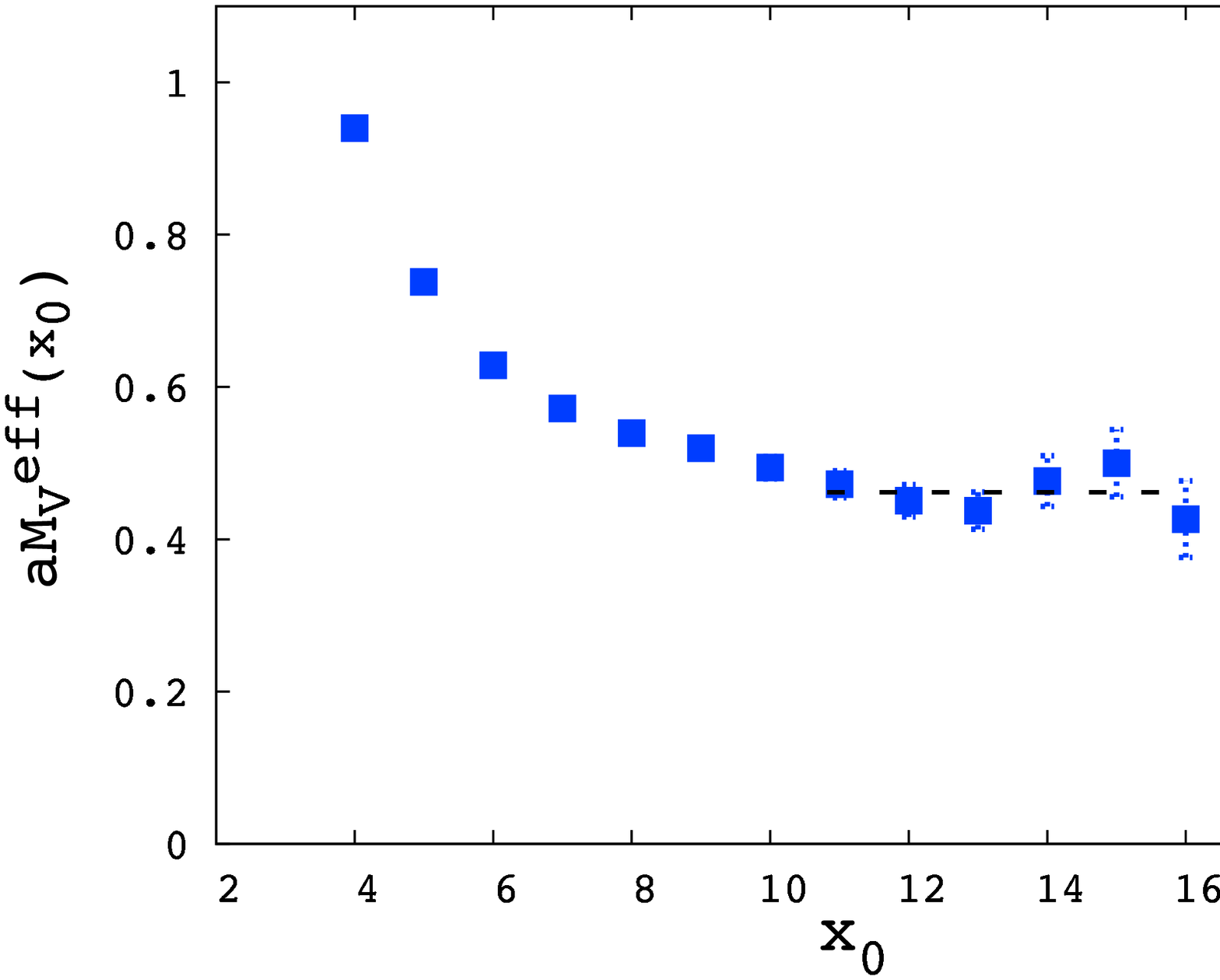}}
\subfigure[]{\includegraphics[scale=0.26,angle=-0]{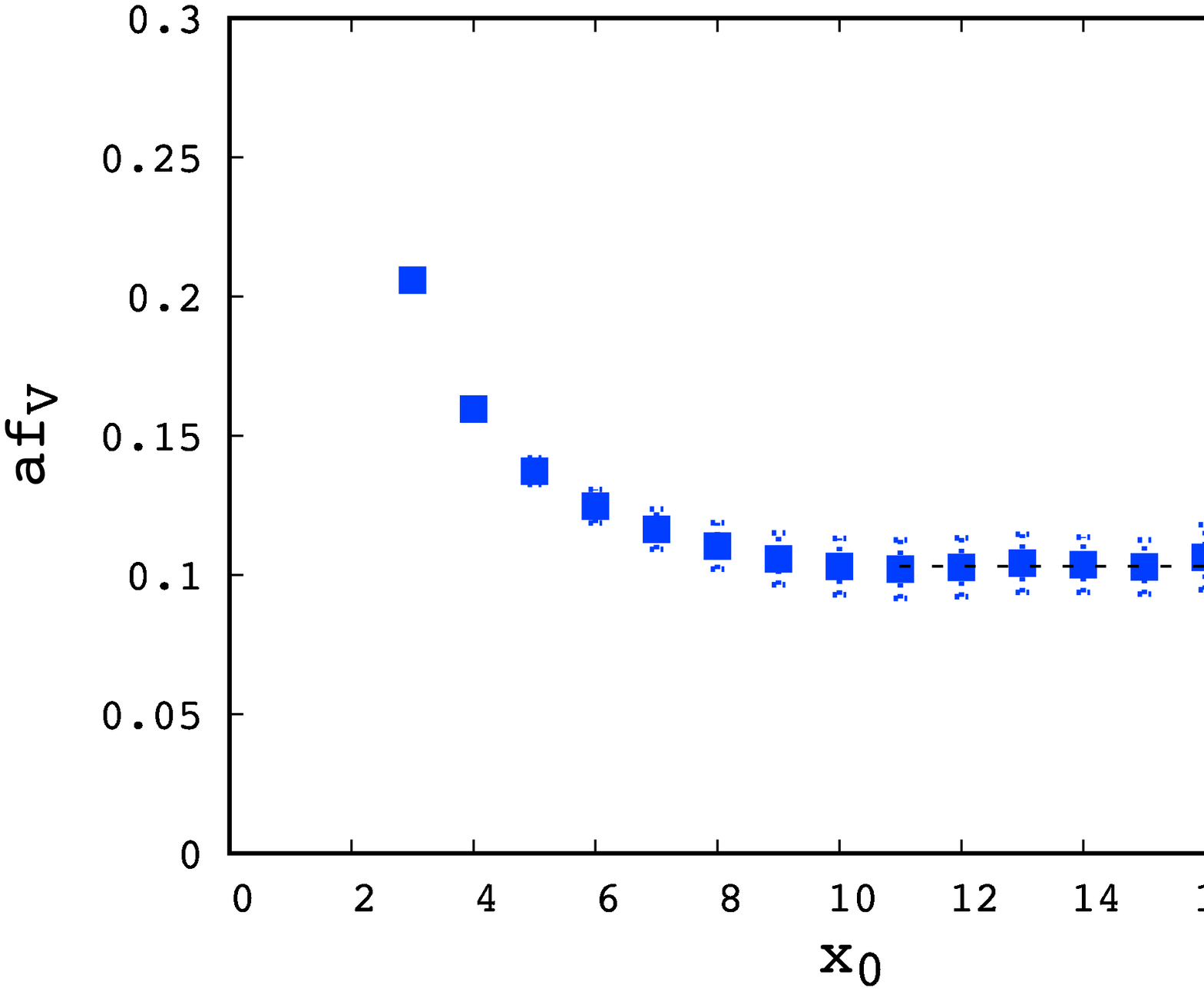}}
\subfigure[]{\includegraphics[scale=0.26,angle=-0]{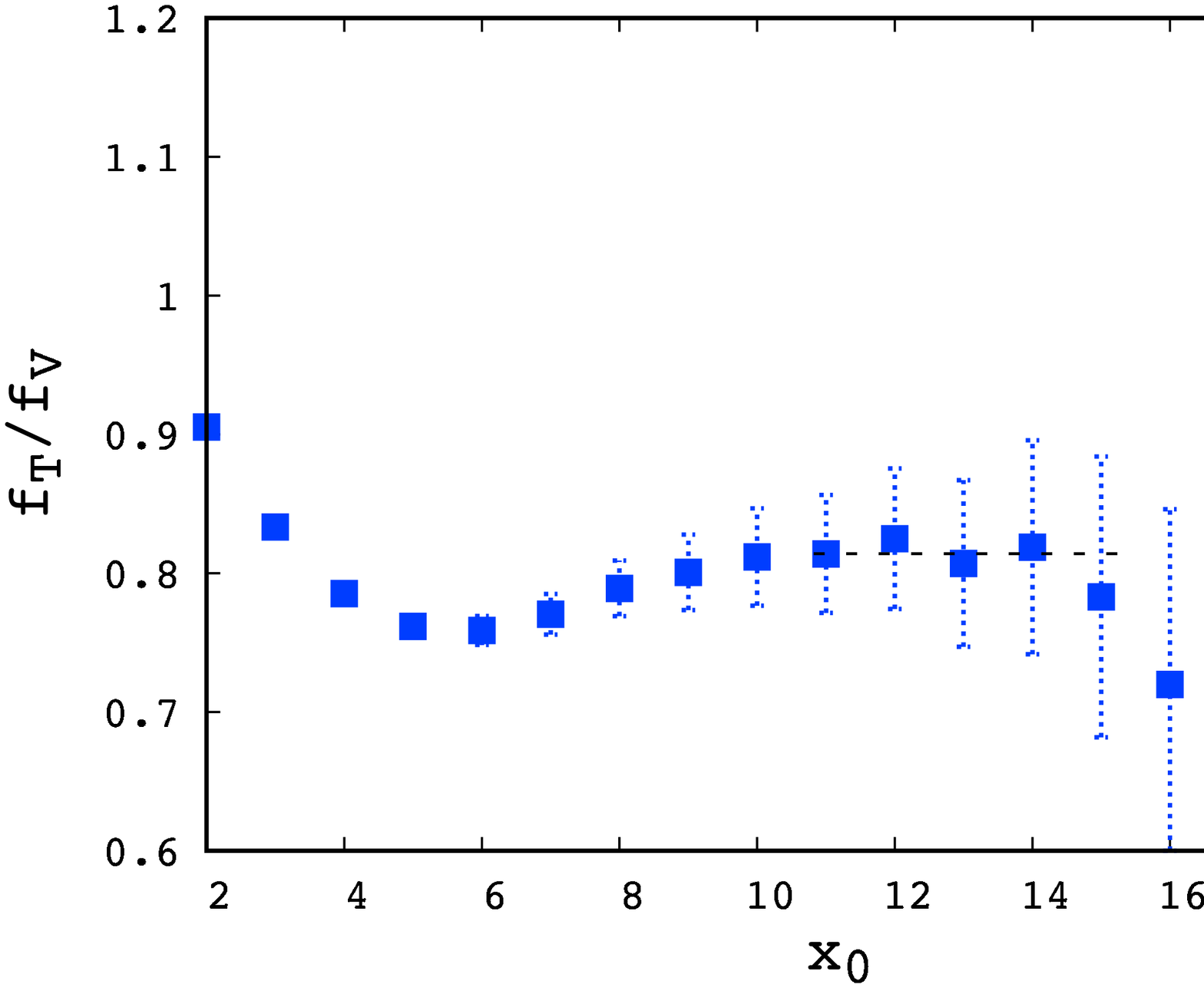}}
\caption[]{
Vector meson quantities plotted as a function of the operators' time separation, computed at $a \mu_l= a\mu_{sea} = 0.0040$ and $a \mu_h = 0.0150$. (a) effective mass; (b) decay constant; (c) ratio $f_T / f_V$.
The quoted results are obtained by taking data over the time plateau $11 \le x_0 \le 16$.
}
\label{fig:plateaux}
\end{center}
\end{figure}

We thus proceed as follows: the strange vector meson consists of two valence quarks $(\mu_l, \mu_h)$.  We compute the necessary observables (vector meson mass, vector decay constant and the ratio of the tensor to vector decay constant) for all combinations of $a \mu_l= a\mu_{sea}$ and $a \mu_{h} = 0.0150$, 0.0220, 0.0270. In this way unitarity holds in the light quark sector, while the heavy valence quark mass, in a partially quenched rationale, spans a range around the physical value $\mu_s$. 
Examples of the quality of our signal (for the lightest $a \mu_l= a\mu_{sea}$ masses) are given
in Fig.~\ref{fig:plateaux}.

\begin{figure}[!h]
\begin{center}
\subfigure[]{\includegraphics[scale=0.26,angle=-0]{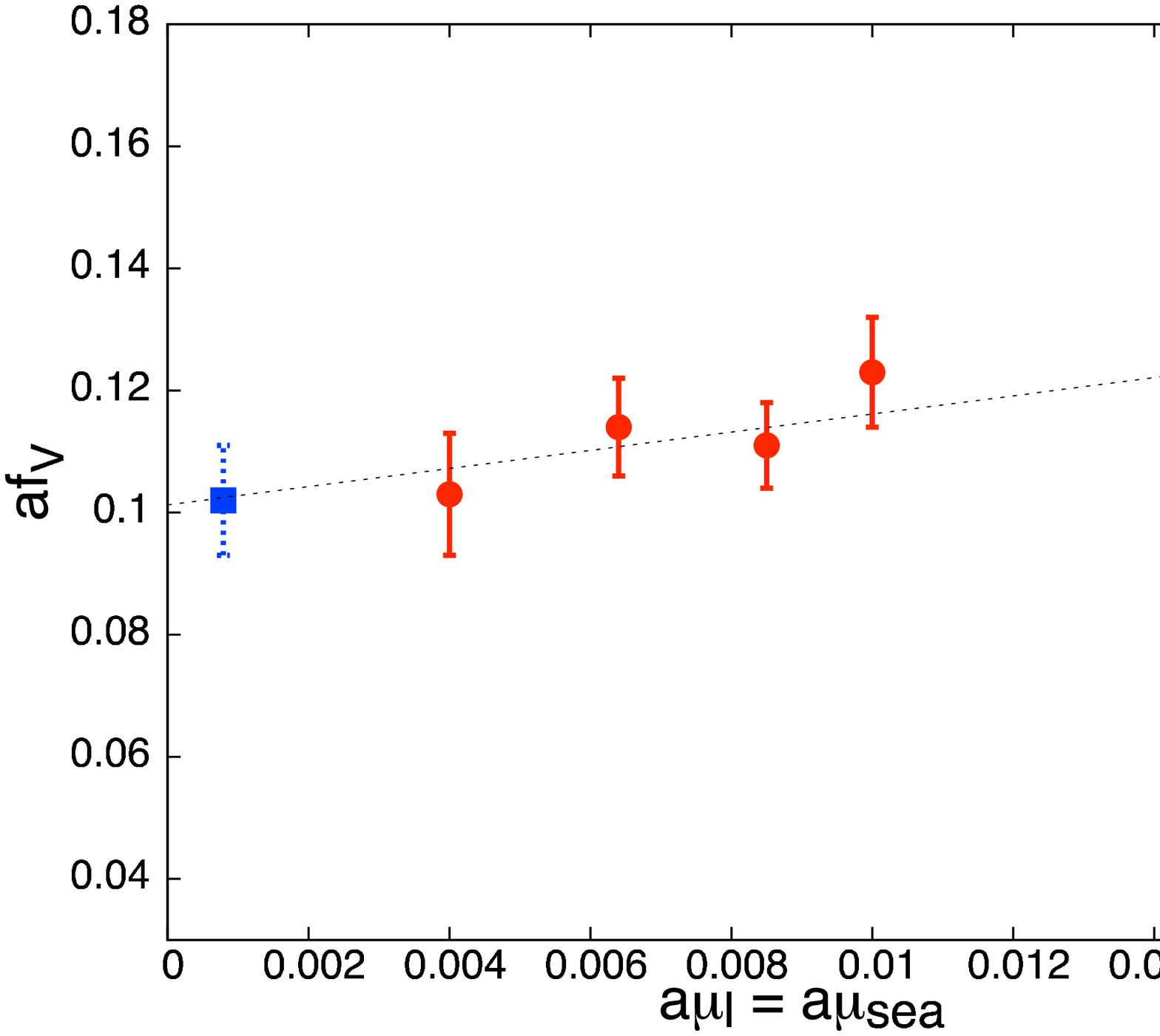}}
\subfigure[]{\includegraphics[scale=0.26,angle=-0]{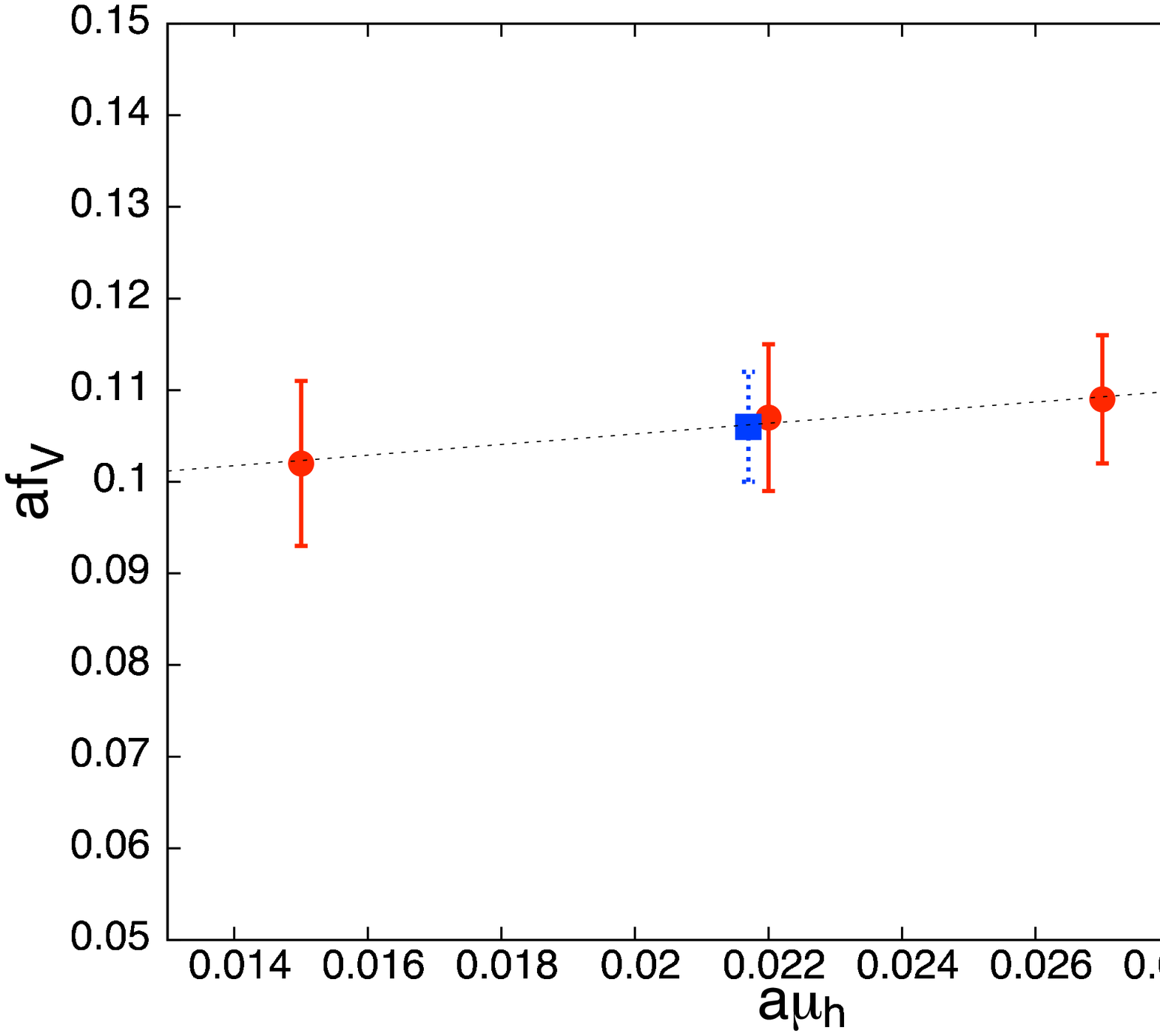}}
\caption[]{
(a) Extrapolation of the vector meson decay constant $f_V$ to the physical down quark mass $a \mu_d$ , at fixed heavy quark mass $a \mu_h = 0.0150$. (b) Interpolation of $f_V(a \mu_d)$ to the physical
strange quark mass $a \mu_s$ of the vector meson decay constant $f_V$. The red circles correspond to
data points, the blue squares to the extrapolation and interpolation results.
}
\label{fig:extrapinterp}
\end{center}
\end{figure}

Subsequently, keeping $a \mu_{h}$ fixed, we extrapolate these results linearly to the physical down quark value $a \mu_d = 0.00079$ (alternative fits are currently under investigation). This is repeated for all three $a \mu_h$ values, before the results are linearly interpolated to the physical strange quark mass $a \mu_s=0.0217(10)$. Examples of the quality of such extrapolations and interpolations are displayed in Fig.~\ref{fig:extrapinterp}.
Our final results are (tmQCD regularization on the lhs and OS regularization on the rhs)
\begin{eqnarray}
aM_V|_{K^{*}} &= 0.422(10)(03) \qquad aM_V|_{K^{*}} &= 0.437(08)(04)
 \nonumber  \\
af_V|_{K^{*}} &= 0.106(05)(02) \qquad af_V|_{K^{*}} &= 0.117(03)(01) 
 \\
f_T/f_V|_{K^{*}} &= 0.770(20)(09) \qquad f_T/f_V|_{K^{*}} &= 0.759(19)(07) 
\nonumber
\end{eqnarray} 
The first error is statistical. The second is systematic and includes the uncertainty of the $a\mu_s$ estimate plus (for the decay constants) that of the renormalization factors $Z_V,Z_A,Z_T$.
The fairly good agreement between tmQCD and OS estimates suggests that cutoff effects in the vector channel are small, in accordance with the expectations of ref.~\cite{FR}. This result must clearly be confirmed by repeating the analysis at other $\beta$ values. The experimental values of the vector meson mass
and decay constants, expressed in lattice units  (at $\beta = 3.9$, we use $a \simeq 0.086~{\rm fm}$)
are $a m_{K^*} = 0.381$ and $a f_{K^*} = 0.0927$, which are reasonably close to our lattice estimates. Finally, the tensor to vector decay constant ratio is RG-run from the scale $a^{-1} \sim 2.3$GeV to the usual reference scale of 2GeV, obtaining $f_T/f_V|_{K^{*}}$ = 0.764(19)(03). This is pretty close to the continuum limit quenched result $f_T/f_V|_{K^{*}}$ = 0.74(2) of ref.~\cite{roma}.

\section{The K-meson bag parameter}

We now apply the proposal of ref.~\cite{Frezz-Rossi2} in an $O(a)$-improved calculation of
$B_{\rm K}$. Recall that, since we require both automatic improvement and multiplicative renormalization of this quantity, the setup is ``mixed": the two Kaon valence quarks are maximally twisted in standard tmQCD fashion,
while the two anti-Kaon ones are OS valence quarks (or vice versa). In this way, the continuum (renormalized) four-fermion operator is related to the lattice bare one as follows (the notation should be obvious)
\begin{equation}
{\cal V}_\mu {\cal V}_\mu + {\cal A}_\mu {\cal A}_\mu \,\, = \,\, Z_{VA+AV} \,\, \Big [
V_\mu^{\rm tm} A_\mu^{\rm OS} \,\, + \,\, A_\mu^{\rm tm} V_\mu^{\rm OS}  \Big ]
\end{equation}

With $x_0 = w$ an arbitrary reference time-slice, the relevant 3-point correlation function consists of two ``K-meson walls" with noise sources at fixed times ($w$ and $w+T/2$) and a 4-fermion operator living at different time-slices $x_0$, with $w < x_0 < w + T/2$. The signal quality is improved in various ways. For example, the reference time $w$ is varied from configuration to configuration. Moreover, as already stated previously, the valence quarks emanating from  $w$ are of the OS
variety, while those emanating from  $w+T/2$ are standard tmQCD ones. This situation is also
reversed and the corresponding correlation functions are suitably averaged.

\begin{figure}[!h]
\begin{center}
\subfigure[]{\includegraphics[scale=0.26,angle=-0]{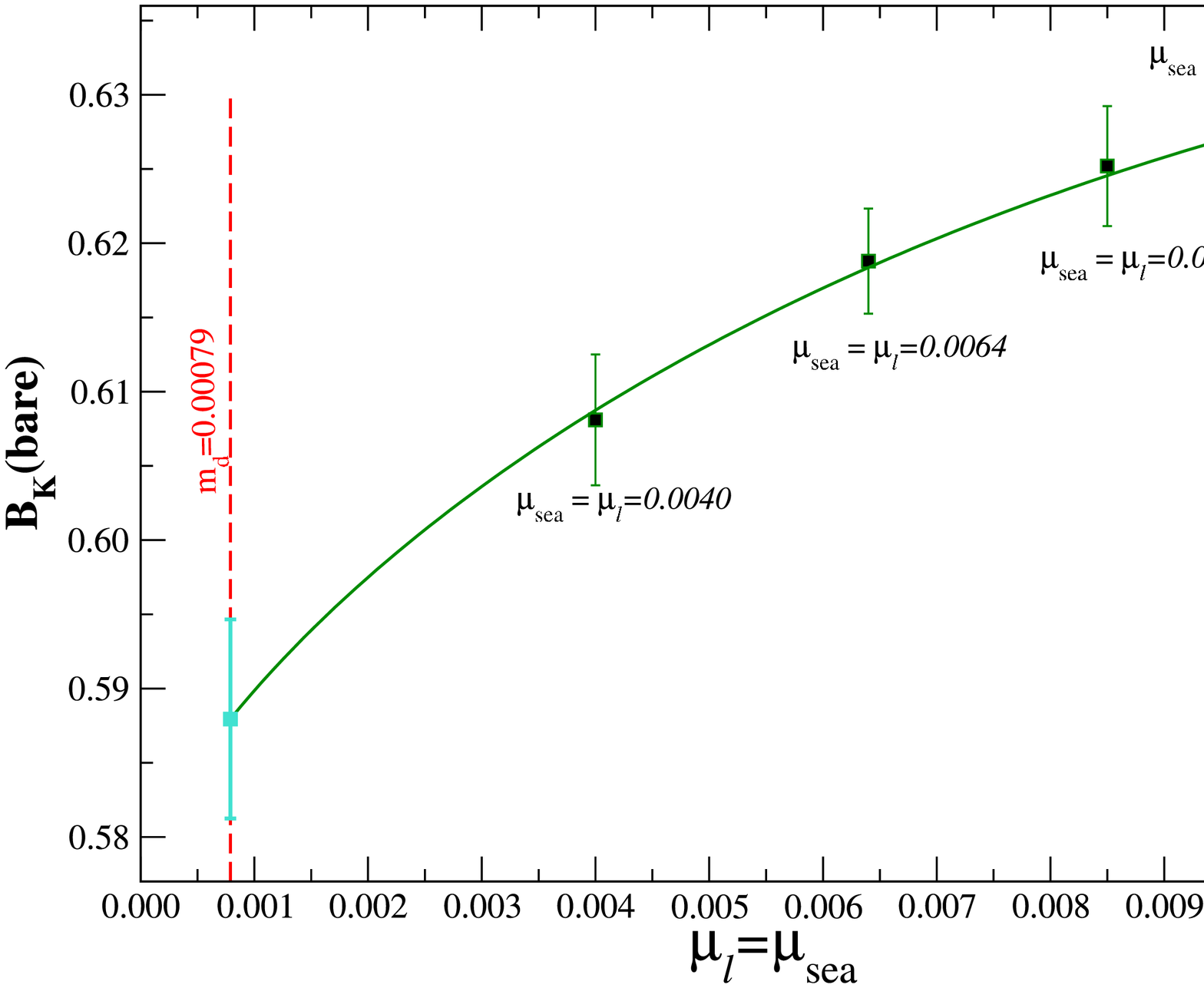}}
\subfigure[]{\includegraphics[scale=0.26,angle=-0]{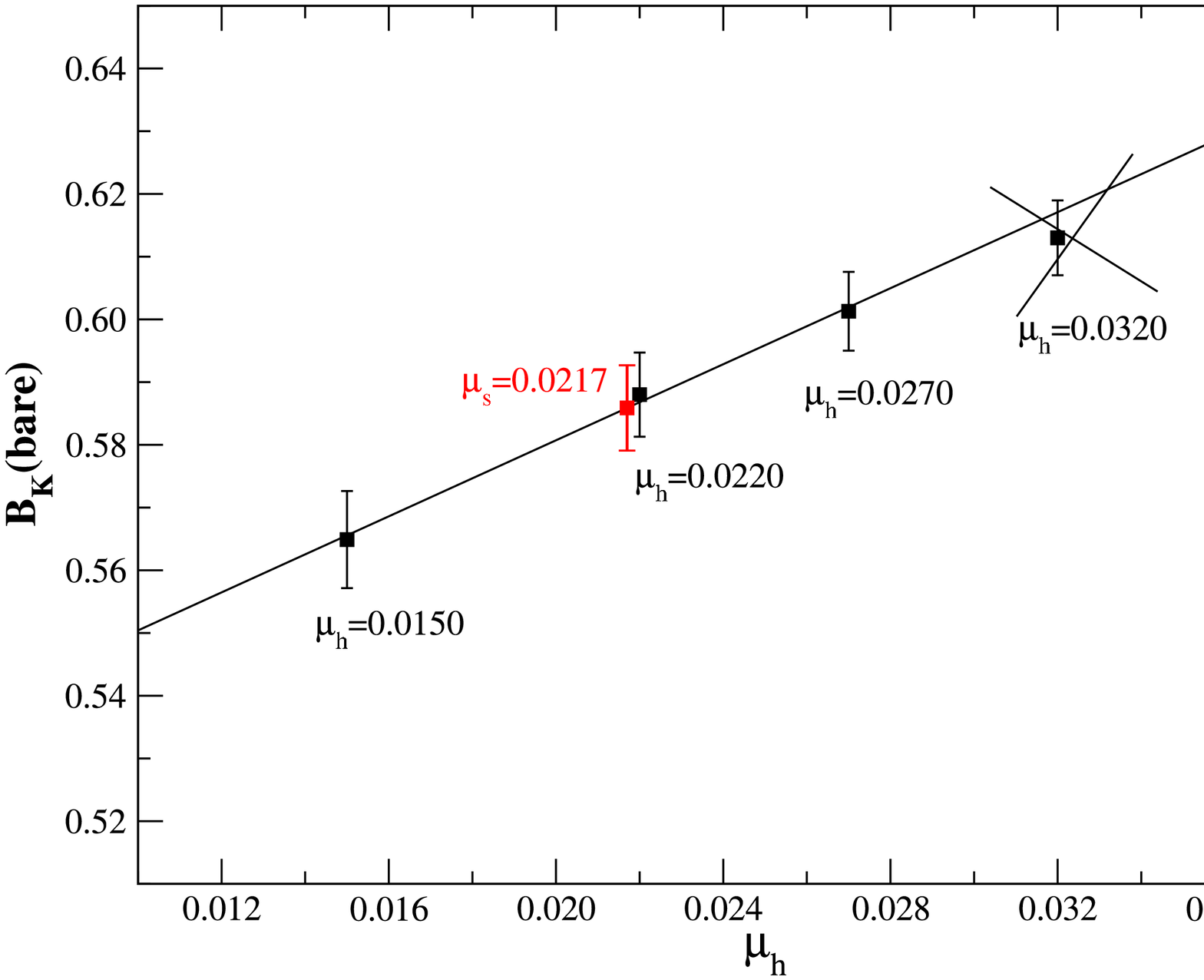}}
\caption[]{
(a) Extrapolation of $B_{\rm K}$ to the physical down quark mass $a \mu_d$, at fixed heavy quark mass  $a\mu_h = 0.0220$. The $B_{\rm K}$ datapoints are computed at degenerate light and sea quark masses, $\mu_l = \mu_{\rm sea}$ and fitted to eq.~(\ref{eq:pqchipt}). (b) Linear interpolation of $B_{\rm K}$, as a function of the heavy valence quark mass $a \mu_h$,  to the physical strange quark mass $a \mu_s$.}
\label{fig:BK}
\end{center}
\end{figure}

The $B_{\rm K}$-parameter is measured from correlation functions with the light valence quark mass kept equal to the sea quark mass ($a\mu_l = a\mu_{\rm sea}$).
At fixed heavy valence quark mass $a \mu_h$, we fit the light mass behaviour in $a \mu_l$, using the $SU(2)$ Partially Quenched Chiral Perturbation Theory (PQ-$\chi$PT) result of refs.~\cite{SharpeZhang,Alltonetal}:
\begin{equation}
B(\mu_h) \,\, = \,\, B_\chi(\mu_h) \,
\Big [ 1 \, + \,  b(\mu_h) \, \frac{2B_0}{f^2} \, \mu_l \, - \, \frac{2B_0}{32\pi^2 f^2} \mu_l \, 
\ln\big (\frac{2B_0\mu_l}{\Lambda_\chi^2} \big ) \Big ]
\label{eq:pqchipt}
\end{equation}
With the coefficient $2B_0/f^2$ known from earlier chiral fits of the light quark
sector~\cite{etmc-light,etmc-strange},
the above relation requires a two-parameter fit ($B_\chi$ and $b(\mu_h)$)
in the chiral region. In this way we extrapolate $B(\mu_h)$ at the physical down quark mass $a \mu_d$. The result is shown in Fig.~\ref{fig:BK}(a). Alternative polynomial fits are currently under study.
The final step is to linearly interpolate the $B(\mu_h)$ estimates to the physical strange quark mass value $a\mu_s$. This is shown in Fig.~\ref{fig:BK}(b). Note that the result of the highest $a\mu_h$ has not been included, as it lies rather far from $a\mu_s$.

Finite size effects appear to be under control, since at  $\mu_{sea} = 0.0040$ and
$\mu_l = \mu_h =0.0100$, we find $B_K^{\rm bare} = 0.591(5)$ for the $L = 24$ lattice and
$B_K^{\rm bare} = 0.598(8)$ for the $L = 32$ one.

\begin{figure}[!h]
\begin{center}
\subfigure[]{\includegraphics[scale=0.26,angle=-90]{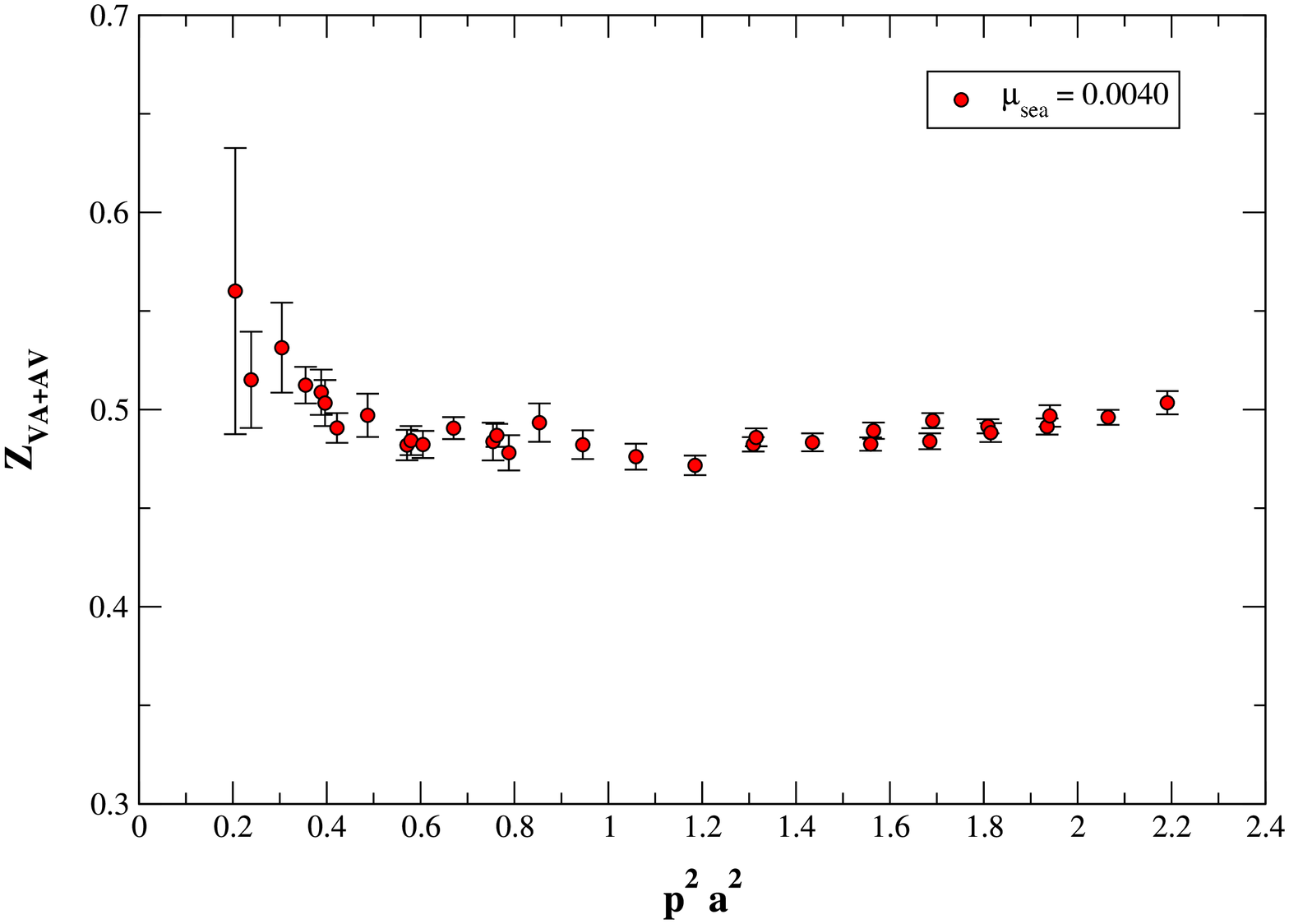}}
\subfigure[]{\includegraphics[scale=0.26,angle=-90]{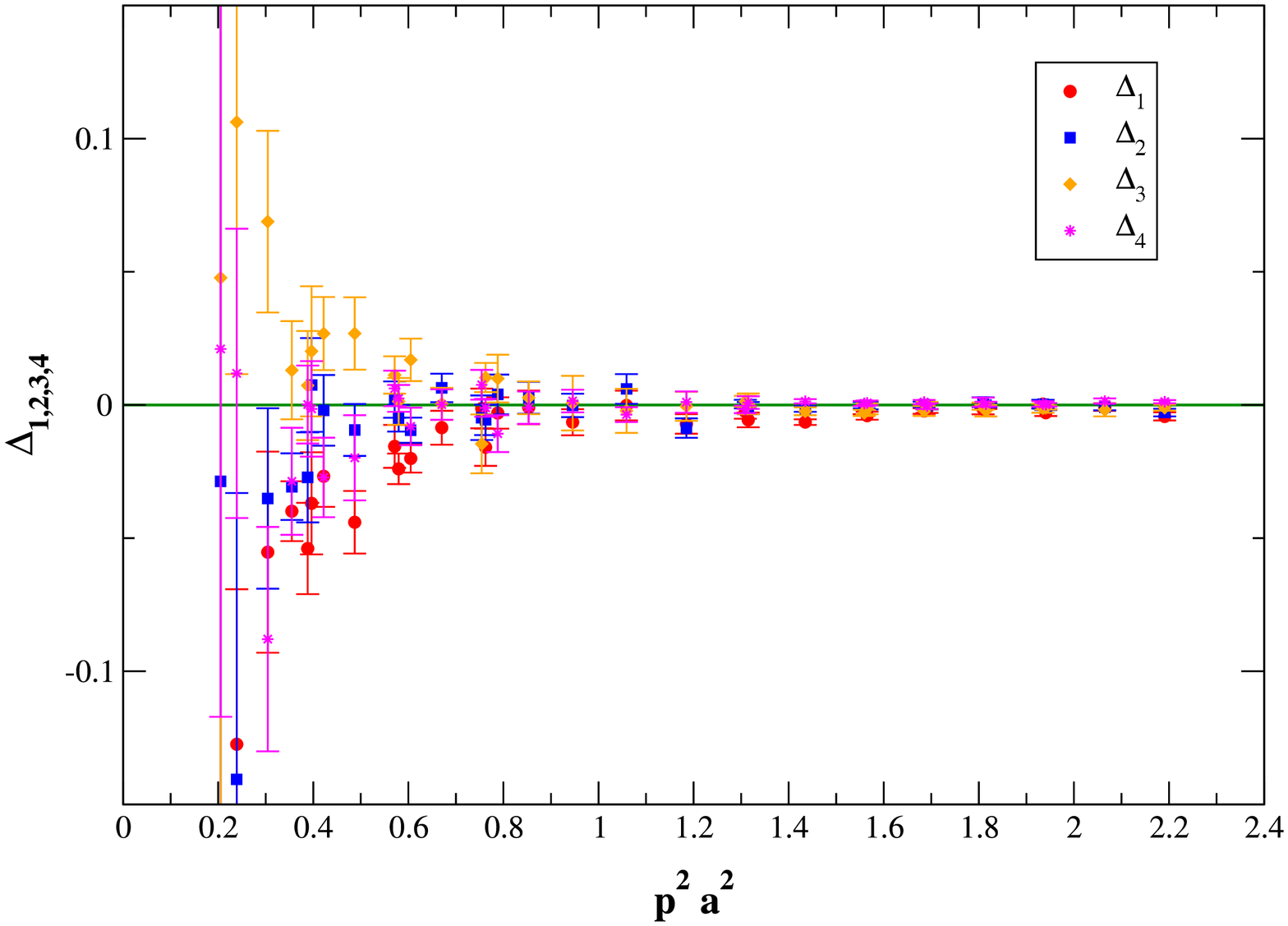}}
\caption[]{(a) RI/MOM computation of the multiplicative renormalization factor $Z_{VA+AV}$.
(b) Mixing coefficients $\Delta_k$ ($k=1,\cdots ,4$) with other four-fermion operators with ``wrong chirality".
}
\label{fig:Zs}
\end{center}
\end{figure}

Renormalization is again performed in the RI/MOM scheme~\cite{rimom4}. The quality of our preliminary results for the multiplicative renormalization factor $Z_{VA+AV}$ is shown in Fig.~\ref{fig:Zs}(a). The absense of mixing
with ``wrong chirality" operators is explicitly demonstrated in Fig.~\ref{fig:Zs}(b), where all mixing
coefficients are shown to be vanishing. We estimate $Z_{VA+AV}(\beta = 3.9; 2{\rm GeV}) = 0.454(18)$.

The results reported in this work are very encouraging, but a word of caution is in place here. At fixed $\beta$, the two Kaon states, obtained with different regularizations (i.e. tmQCD and OS) are not degenerate, differing by $O(a^2)$ discretization terms. The two different exponential decays, as well 
as the factors $m_{\rm K}^{\rm tm}$ and $m_{\rm K}^{\rm OS}$ of the matrix element $<~\bar K^0 (m_{\rm K}^{\rm tm}) | O_{VA+AV} | K^0 (m_{\rm K}^{\rm OS}) > \propto~m_{\rm K}^{\rm tm} m_{\rm K}^{\rm OS}$, cancel out in the ratio of $B_{\rm K}$.
The Kaon mass splitting is sometimes quite significant; for the $\beta = 3.90$ case in hand, with $a\mu_l = a\mu_{\rm sea} =  0.0040$ and $a\mu_h = 0.0220$ we have $m_{\rm K}^{\rm tm} = 0.2391(07)$ and 
$m_{\rm K}^{\rm OS} = 0.2923(16)$. It is important to monitor the size of this splitting with increasing $\beta$, in order to confirm that it vanishes like $a^2$.

Our preliminary result at a single lattice spacing is $B_{\rm K}(2 {\rm GeV}; {\rm RI/MOM}) = 0.56(2)$, corresponding to $B_{\rm K}^{\rm RGI} = 0.77(3)$.

\section*{Acknowledgements}
We thank our ETM collaborators for their help and encouragement. This work has been supported in part by the EU ITN contract MRTN-CT-2006-035482, ``FLAVIAnet".

\end{document}